\documentclass[prb,twocolumn,superscriptaddress]{revtex4}
\usepackage{amsmath}
\usepackage{amssymb}
\usepackage{wasysym}
\usepackage{graphicx}
\usepackage{times}
\usepackage{longtable}

\begin{document}

%%%%%%%%%%%%

\title{Gauge symmetry and non-abelian topological sectors\\ in a
  geometrically constrained model on the honeycomb lattice}

\author{Paul Fendley$^1$, Joel E. Moore$^{2,3}$, and Cenke Xu$^2$
\medskip \\ 
$^1$ Department of Physics, University of Virginia, \\
Charlottesville, VA 22904-4714 USA 
\smallskip\\
$^2$  Department of Physics, University of California, \\
Berkeley, CA 94720
\smallskip\\
$^3$ Materials Sciences Division, Lawrence Berkeley National Laboratory, \\
Berkeley, CA 94720
}

\date{January 17, 2007}

%\maketitle

\begin{abstract}
We study a constrained statistical-mechanical model in two dimensions
that has three useful descriptions. They are 1) the Ising model on the
honeycomb lattice, constrained to have three up spins and three down
spins on every hexagon, 2) the three-color/fully-packed-loop model on
the links of the honeycomb lattice, with loops around a single hexagon
forbidden, and 3) three Ising models on interleaved triangular
lattices, with domain walls of the different Ising
models not allowed to cross.  Unlike the three-color model, the configuration
space on the sphere or plane is connected under local moves. On
higher-genus surfaces there are infinitely many dynamical sectors, 
labeled by a noncontractible set of nonintersecting
loops.  We demonstrate that at infinite temperature the transfer
matrix admits an unusual structure related to a gauge symmetry for the
same model on an anisotropic lattice. This enables us to
diagonalize the original transfer matrix for up to 36 sites, finding
an entropy per plaquette $S/k_B \approx 0.3661\dots$ and substantial
evidence that the model is not critical. We also find the striking
property that the eigenvalues of the transfer matrix on an anisotropic
lattice are given in terms of Fibonacci numbers.  We comment on the
possibility of a topological phase, with infinite topological
degeneracy, in an associated two-dimensional quantum model.

\end{abstract}

\maketitle

\section{Introduction}

Classical lattice statistical-mechanical models with local constraints
have been of great interest for decades. By ``local constraint'', we
mean a local rule which restricts the allowed configurations. A famous
example is that of the hard-core close-packed dimer model
\cite{Kastelyn,Fisher63}. The degrees of freedom are dimers stretching
between adjacent sites of a lattice, while the hard-core and
close-packing constraints mean that each site of the lattice is
touched by exactly one dimer. Another famous example is Baxter's
three-color model, where each link is covered by one of three
``colors'' of dimers, with the constraint that each site is touched by
all three colors \cite{Baxter70}.

Another oft-studied constraint is to require that the degrees of freedom be
``loops'', i.e.\ one-dimensional objects
without ends. For example, both close-packed hard-core dimers and the
three-color model can be viewed as loop models. In the latter case,
the links colored by two of the colors (say $R$ and $G$) form closed
loops of alternating $R$ and $G$ colors. Since every vertex has one
$R$ and one $G$ touching, the three-color model is therefore
equivalent to a fully-packed loop model (every site has one loop going
through it). Each loop receives a weight $2$, since there are two
possible ways of ordering $R$ and $G$ around each loop.

One interesting limit of constrained models is at infinite
temperature, where each allowed configuration has the same Boltzmann
weight. The partition function in this limit is a purely combinatorial
quantity: it simply counts the number of configurations. Because of
the constraints, the physics of such models is still very rich. For
example, the three-color model is critical, as is the hard-core
close-packed dimer model on the square lattice
\cite{Fisher63}. Obviously, not all constrained models are critical:
dimers on the triangular lattice (or any non-bipartite lattice) have
exponentially decaying correlators \cite{Moessner01}.

The purpose of this paper is to present a constrained lattice model
that has several rather interesting properties.  There are three
equivalent ways of defining the model. One is as an Ising model on the
honeycomb lattice with a constraint around each hexagon; one is with a
constraint on the three-color model, and a third is as three coupled
Ising models. The latter form is most naturally given in terms of
loops representing Ising domain walls, and is also the representation
where its properties are most transparent.

This model is of interest for several reasons. It is defined in terms
of simple local degrees of freedom and constraints, yet exhibits
fascinating conservation laws. As we will detail, the transfer matrix
decomposes into sectors which are labeled by non-abelian (and
non-local) charges. The number of distinct sectors {\em exponentially}
increases as the size of the system increases. This symmetry enables
us to do exact diagonalization of the transfer matrix for systems of
sizes up to 36 sites across, i.e.\ a Hilbert space initially of size
$2^{36}$. We know of no non-trivial system with such a property.

Despite the fact that the conserved charges are non-local, the
configuration space of the model has the striking property that it is
connected under simple local moves. Even though the three-color model
is closely related to ours, to relate all its different configurations
requires changing degrees of freedom arbitrarily far apart
\cite{henley,huserutenberg}. Since we give a precise relation between
the three-color model and ours, we thus have located the obstruction
to connectivity (the ``11th'' vertex discussed below) in the
three-color model. Not only does this mean that our model is amenable
to Monte Carlo situations, but it should prove interesting to study
its classical dynamics \cite{henley,chamon2}.

A new reason to be interested in two-dimensional classical lattice
models with constraints comes from {\em quantum} physics. The
motivation is to find phases with topological order, where there is no
non-vanishing local order parameter, but only non-local ones. The idea
for building such a quantum model by starting with a classical magnet
with local constraints came long ago \cite{andersonfazekas}, and a
theoretical triumph in proving they exist came from a quantum
eight-vertex model \cite{Kitaev97} and a quantum dimer model on the
triangular lattice \cite{Moessner01}. The two-dimensional quantum
models are defined by using each configuration in a two-dimensional
classical lattice model as a basis element of the
Hilbert space. One characteristic of topological order is that the
number of ground states depends on the genus of two-dimensional
space. Constrained lattice models give natural ways of defining the
different sectors which, with appropriate choice of Hamiltonian
\cite{Rokhsar88}, correspond to different ground states in the quantum
theory. When writing the eight-vertex or dimer models as loop models, the
different ground states are labeled by the number (mod 2) of loops
which wrap around cycles of the torus.

In section \ref{sec:models} we introduce these models and show that
they are equivalent. We relate our model to several others in appendix
\ref{sec:relations}, enabling us to put upper and lower bounds on the
entropy. In section \ref{sec:flip}, we discuss the dynamics under
local moves, showing that the configurations on the plane or sphere
are all connected by simple local moves. On surfaces with
non-contractible cycles, we classify the infinitely many separate dynamical
sectors. In section \ref{sec:ft}, we give arguments which suggest
our model is not critical.  We also develop in section
\ref{sec:transfer} and solve in section \ref{sec:gauge} a
closely related model which has a gauge symmetry. We exploit this
symmetry in section \ref{sec:numerics} to show how to reduce
dramatically the size of the original transfer matrix. This enables us
to exactly diagonalize the transfer matrix for quite large lattices,
and the results again suggest that the model is not critical.  In
section 9, we present our conclusions, and discuss applying our
results to build a quantum model with topological order.

\section{The model, and its three descriptions}
\label{sec:models}

The model we are introducing can be described in three equivalent
ways. Here we present them, and then demonstrate their equivalence.

\paragraph{Model 1:}

The degrees of freedom of the Ising model are ``spins'' $\sigma_i$
taking values of $\pm 1$ at each site $i$ of some lattice. The energy
in general from nearest-neighbor interactions is given by
$$
E= J \sum_{\langle ij\rangle} \sigma_i\sigma_j. 
$$
The Ising model on the honeycomb lattice has a critical
point when $K=J/k_BT= \hbox{arcsinh}(\sqrt{3})/2$\ \cite{Mccoybook}.
Our model 1 is the Ising model on the honeycomb lattice, with the
constraint that there must be three up spins and three down spins around
each hexagon, i.e.\ 
\begin{equation}
{\bf model\ 1:}\qquad\qquad
M_h\equiv \sum_{i\in \hexagon} \sigma_i = 0 
\label{constraint1}
\end{equation}
This is quite a strong constraint, retaining only 20 of the original
64 possibilities for the spins around each hexagon. We will mostly
discuss the infinite temperature limit
$T\to\infty$ or $K=0$, in which each allowed configuration has equal
weight.

\paragraph{Model 2:}

The degrees of freedom in the three-color model are three colors, say
$R$, $G$, and $B$, which are placed on the links on the honeycomb
lattice. The usual constraint in the three-color model is to require
that at each site of the lattice, all three colors appear. In other
words, links of the same color can never touch. When the partition
function is simply the sum over all allowed configurations (i.e.\ in
the infinite-temperature limit), the model is critical and integrable 
\cite{Baxter70}. Our model 2 is the
three-color model with an additional constraint forbidding configurations 
which have the same two colors alternating around any given hexagon. In
a picture,
\begin{equation}
{\bf model\ 2:}\qquad {\bf forbid} \qquad
\begin{picture}(90,35)
\put(25,-29){$c$}\put(0,-12){$c'$}\put(0,12){$c$}\put(25,29){$c'$}
\put(50,-12){$c'$}\put(50,12){$c$}
\end{picture}
\label{constraint2}
\end{equation}

\bigskip\bigskip\medskip
\noindent
where $c\ne c'$ can be any of $R$, $G$, or $B$. This forbids 6 of the 66
allowed configurations around a hexagon in the three-color model. 

Imposing the constraint (\ref{constraint2}) in the fully-packed loop
formulation of the three-color model forbids the shortest loops, of
length 6. The constraint is symmetric under permutations of $R$, $G$,
and $B$, so it forbids all ``short'' loops, no matter which two colors
are chosen to form the loops.

\begin{figure}[h]
\centerline{\includegraphics[width=.5\textwidth]{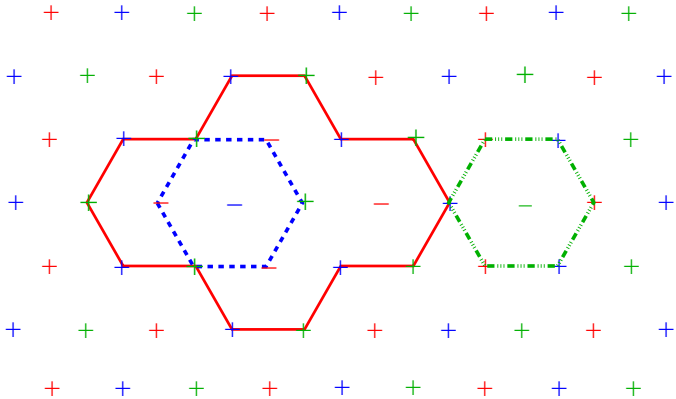}}
\caption{Three Ising models on three triangular sublattices. The
  constraint (\ref{constraint3}) requires that the domain walls do not
  cross.} 
\label{fig:3Ising}
\end{figure}

\paragraph{Model 3:}
Consider now Ising spins $s_i=\pm 1$ on the triangular
lattice. Instead of studying a single Ising model on this lattice, we
instead consider {\em three} identical Ising models, on each of the
three identical triangular sublattices of the triangular lattice. This
has an (Ising$)^3$ critical point when
$K=\hbox{arcsinh}(1/\sqrt{3})/2$\ \cite{Mccoybook}. The domain walls for
an Ising model separate unlike spins; they live on the links of the
dual lattice.  For each of our three Ising models on triangular
lattice, its dual lattice is the honeycomb lattice made up of the
sites of the other two Ising models. It is not possible for the domain
walls of a given Ising model to cross or even touch, but the walls of
the different decoupled models can cross and touch. Our constraint
couples the three Ising models by not allowing the walls of different
models to cross (although they can touch). A configuration in this
model is displayed in figure \ref{fig:3Ising}. In terms of the spins,
consider a hexagon on the triangular lattice, comprised of six sites
surrounding a given site. Label the six spins around this hexagon by
$s_i$, so that $s_2$, $s_4$ and $s_6$ are in one of the three Ising
models, while $s_1$, $s_3$ and $s_5$ are in another. Denote
$d_i=s_is_{i+2}$, with the subscripts interpreted mod $6$. A domain
wall occurs when $d_i=-1$. The constraint that domain walls not
cross in terms of these spins is then
\begin{equation}
{\bf model\ 3:}\qquad
{\cal C}_j\equiv 3-\sum_{i=1}^6 (d_i-d_id_{i+1}+d_id_{i+3}/2) =0 
\label{constraint3}
\end{equation}
where $j$ is the site on the triangular lattice at the center of the hexagon.
 The allowed domain walls inside this hexagon
are of the types illustrated in figure
\ref{fig:10vertex2} below. 
Model 3 can equivalently be described in terms of closed
mutually-avoiding loops on the triangular lattice, with the
added restriction that loops must turn by $\pm 120$ degrees at every site.

\bigskip

These three models are equivalent to each other under local
reformulations of the degrees of freedom.  First let us recall the
mapping of the three-color model without constraint
(\ref{constraint2}) to an Ising model on the honeycomb lattice
\cite{difrancesco94}.  The Ising variables represent chiralities in
the three-color model; this chirality representation occurs in the
superconducting-array realization of the three-color
model~\cite{moorelee,chamon}.  Consider  given configuration in
the three-color model.  There are six possible configurations of the
three-color model around each site of the honeycomb lattice: 
%the three
%links with ending in this site can be colored $RGB$ going clockwise,
%or $RGB$ going counter-clockwise.
% Divide the sites of the
%honeycomb lattice into two equivalent sublattices. 
%For the sites on sublattice 1, 
Put an Ising spin $+$ on the site if the colors on the three links
touching it are $RGB$ clockwise, and $-$ if the three are $RGB$
counter-clockwise.
%On sublattice 2, do the opposite. 
Going around each
hexagon, it is easy to check that there are either $0$, $3$ or $6$ up
spins. It is also easy to check for any configuration with $0$, $3$ or
$6$ up spins, one can reverse the map and find a configuration in the
three-color model. Ignoring boundary conditions, there are three
configurations in the three-color model for each in the Ising model,
so the map can be made one-to-one by specifying the color on one
link. The
three-color model at infinite temperature therefore maps onto the
Ising model on the honeycomb lattice with $K=0$ and the requirement
the sum of the $\sigma_i$ around hexagon obeys $M_h= 0,\pm 6$. One can
generalize the three-color model to include interactions equivalent to
a non-zero $K$ if desired; this is easily done in the domain-wall
formulation given below.

The equivalence of model 1 to model 2 is now obvious. Hexagons in the
three-color model with alternating colors as in (\ref{constraint2})
correspond to having $M_h=\pm 6$ in the Ising model. These are forbidden
in model 1 by (\ref{constraint1}), and in model 2 by
(\ref{constraint2}). 

\begin{figure}[h]
\centerline{\includegraphics[width=.46\textwidth]{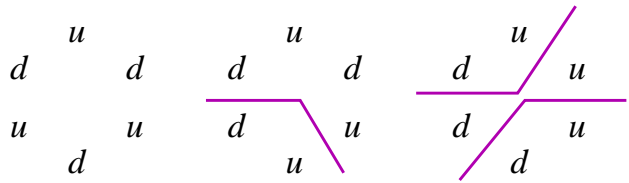}}
\caption{The three types of antidomain-wall configurations on a
  hexagon in model 1. The Ising spins here are denoted by $u$ and
  $d$ to distinguish them from the Ising spins in model 3, which are
  denoted by $\pm$.}
\label{fig:10vertex}
\end{figure}

To show the equivalence of model 1 with model 3, we reexpress the
degrees of freedom in model 1 in terms of {\em anti}domain walls.
Every time adjacent spins are different, we draw an antidomain wall on
the link of the dual lattice separating them. These antidomain walls
therefore form loops on the dual triangular lattice. Each of the
configurations on each hexagon obeying the constraint
(\ref{constraint1}) correspond to one of the types of antidomain-wall
configurations illustrated in figure \ref{fig:10vertex}.
There are 10 different configurations of three different types: the
empty one, six (related by 60 degree
rotations) with two antidomain walls, and three (related by 60 degree
rotations) with four antidomain walls.  Ising domain walls have a
weight $e^{-2K}$ per link. Since antidomain walls are simply the
complement of the domain walls, they can be taken to have weight
$e^{2K}$ per link.

These antidomain walls in model 1 correspond to {\em domain} walls in
model 3. The triangular lattice for model 3 is simply the dual lattice
of the honeycomb lattice for model 1. The domain walls in a
triangular-lattice Ising model make a $\pm 120$ degree turn at every
site, just like the antidomain walls in figure \ref{fig:10vertex}.  If
the three Ising models in model 3 were decoupled, there would be 16
different domain-wall configurations going through each site of the
triangular lattice, because there are four possibilities for each of
the two Ising models whose domain walls go through this point.
There are only 10 possibilities in figure \ref{fig:10vertex}. 
\begin{figure}[h]
\centerline{\includegraphics[width=.5\textwidth]{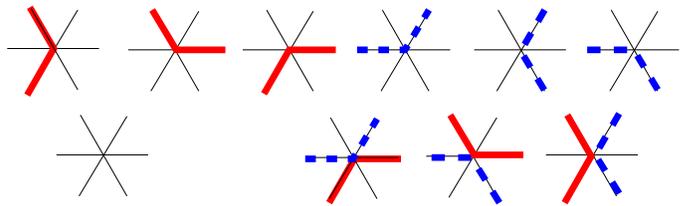}}
\caption{The 10 possible domain-wall configurations in model 3, i.e.\
  the 10 vertices.}
\label{fig:10vertex2}
\end{figure}
Model 1 and
Model 3 are therefore equivalent if we restrict to these 10, which are
redrawn in figure \ref{fig:10vertex2}. As is obvious from the figures,
the ones disallowed are those where the domain walls
cross. Disallowing crossings leaves exactly the 10, so the non-crossing
constraint is the only one. In figure \ref{fig:3Ising}, we drew the
domain walls for the three different Ising models with solid, dashed,
and dot-dashed lines, to emphasize the fact that they do not cross,
with each forming closed loops. In figure \ref{fig:10vertex2}, we drew
these with the dotted and dashed lines, but the same 10 vertices occur
at any point on the triangular lattice with the appropriate types of lines. 
We have therefore shown that model 3 is the
same as model 1, up to unimportant constants
%(3 in front of model 3, 2 in front of model 1) 
in front of the partition functions. Our model is therefore a
``10-vertex model'' on the triangular lattice. These vertices are a
subset of those in the 32-vertex model discussed in \cite{Baxbook}.

\begin{figure}[h]
\centerline{\includegraphics[width=.1\textwidth]{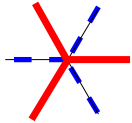}}
\caption{The 11th vertex in the three-color model without constraint
(\ref{constraint2}) }
\label{fig:11thvertex}
\end{figure}

These proofs of course mean that model 2 is equivalent to model 3 as
well, so the three-color model with and without
(\ref{constraint2}) can also be written in terms of a vertex model
on the triangular lattice. The usual three-color model also allows the
11th vertex pictured in figure \ref{fig:11thvertex}.
This vertex is a source/sink of domain walls, and so the three-color
model without constraint (\ref{constraint2}) cannot be mapped onto
three Ising models. 
In model 1, this 11th vertex
corresponds to a hexagon with all up or all down spins, i.e. $M_h=\pm 6$.

In the appendix, we relate our model to two others: hard hexagons on
the triangular lattice, and a generalized Ising antiferromagnet. Our
model is found from these by relaxing constraints, so these models
give lower bounds on the entropy of ours.

\section{Connecting configurations by local moves}
\label{sec:flip}

An important question in many physical applications of two-dimensional
geometrically constrained models is whether the space of states is
connected under local moves.  It is essential if one is to study
either classical or quantum dynamics, and is also useful for doing
Monte Carlo simulations \cite{Haas}.  For example, to build the quantum models
discussed in the introduction, without connectivity under local moves,
the Hamiltonian is non-local. The three-color model is not connected:
any closed loop of bonds of the honeycomb lattice that contains only
two colors will give a different configuration if those two colors are
permuted.  Even though there exist short loops on the lattice (the
shortest loop is a single hexagon), the space of states is not
connected unless the dynamics is able to permute arbitrarily large
loops~\cite{henley,huserutenberg,chamon2}.

In this section we discuss the properties of our model under
local dynamics.  We show that, unlike the three-color model, the
connected sectors can be enumerated simply and correspond to
topological classes of sets of nonintersecting loops in the plane.
Since we have shown that the constraint (\ref{constraint2}) turns the
three-color model into our model, this result illuminates the reason
why the three-color model is not connected by local moves.

\begin{figure}[h]
\centerline{\includegraphics[width=.35\textwidth]{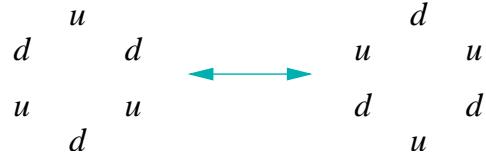}}
\caption{The flip in terms of Ising spins in model 1}
\label{fig:flippable}
\end{figure}

The most-local dynamics of the Ising variables of model 1 that
conserves the constraint (\ref{constraint1}) is to act on hexagons
where the spins alternate up and down around the hexagon.  Flipping
each up spin to down and each down to up around such a hexagon
preserves the constraint not only on the original hexagon, but also on each of
its six neighbors as well. We display this flip in
figure~\ref{fig:flippable}. 
In model 3, this corresponds simply to flipping the Ising spin 
at the center of this hexagon, i.e.\ sending $s_i\to
-s_i$.

\begin{figure}[h]
\centerline{\includegraphics[width=.5\textwidth]{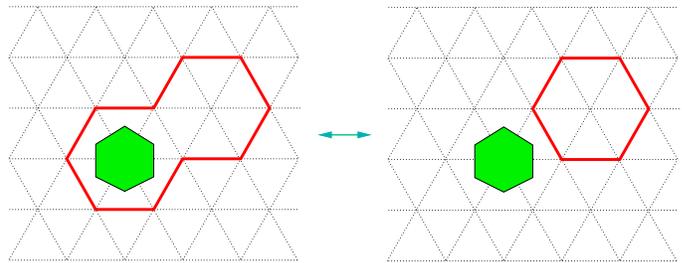}}
\caption{An example of the effect of a flip on a loop. The alternating
  Ising spins in model 1 around the shaded hexagon are those flipped,
  as illustrated in figure \ref{fig:flippable}.}
\label{fig:moveloop}
\end{figure}

This is the only local move necessary to connect configurations. This
is easiest to see in the loop representation.  Since in the model 3, the
flip changes the spin at the center of this hexagon, it simply flips the
model-3 loop variables on the hexagon surrounding this hexagon of
model 1.  An example is illustrated in figure \ref{fig:moveloop}.
For example, if all six of the links on the surrounding hexagon
are empty, the flip creates a
loop of minimal length. If
they are all full, this is a minimal-length loop surrounding the
hexagon, and the flip removes the loop.  In other cases, it shrinks or
expands the loop without creating any loose ends.

It is now easy to see how the flip connects configurations.  A loop of
minimal length has a flippable hexagon inside it, so these can be
removed by one flip. Longer loops can be shrunk and then removed by
repeatedly flipping. If there are loops inside other loops, then the
ones inside need to be removed first.  When space is topologically a
sphere, all configurations are therefore connected to the empty
one. Since all processes can be reversed, this means all
configurations on the sphere are connected.

\begin{figure}[h]
\centerline{\includegraphics[width=.5\textwidth]{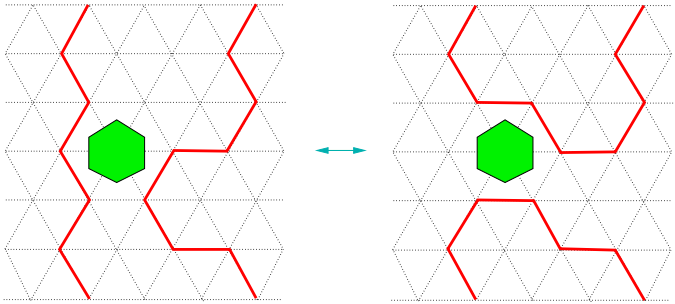}}
\caption{A flip which converts
  non-contractible loops into contractible ones.}
\label{fig:moveloop2}
\end{figure}

When space has non-contractible cycles, however, not all loops can be
removed.  In order to use the formulation of model 3, the periodic
boundary conditions around a cycle must identify sites of the same
triangular sublattice. When this is done, the loops are of three
distinct types, as seen in figure \ref{fig:3Ising}.  Since flips
cannot move two loops of different types through each other, loops
which wrap around a non-contractible cycle can only be removed if they
adjacent to another of the same type. The flip illustrated in figure
\ref{fig:moveloop2} turns two adjacent non-contractible loops of the
same type into two contractible ones.

\begin{figure}[h]
\centerline{\includegraphics[width=.4\textwidth]{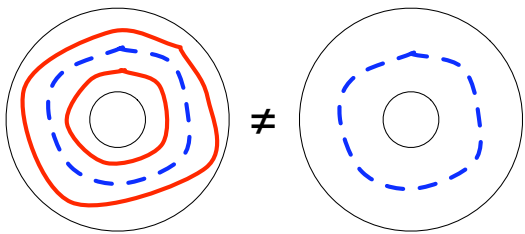}}
\caption{Inequivalent sectors on the annulus, which is topologically
equivalent to a finite cylinder.  The two outer loops cannot move through
the inner loop to annihilate, because of the non-intersection constraint.}
\label{fig:toposector}
\end{figure}

When space is a cylinder, the different 
sectors can be enumerated simply: a sector is given by a sequence of
loop colors (those encountered reading from left to right along the
cylinder, for example), with an even number of adjacent occurrences of
the same colors being equivalent to the identity.  Mathematically,
this set is isomorphic to the free group on three elements $a,b,c$,
with the relations $a^2 = b^2 = c^2 = 1$.  Putting two cylinders next
to each other defines a group action on the set of topological
sectors, and this group action is non-abelian: for example, $a b a b$
is not the same as $a^2 b^2 = 1$.  Fig.~\ref{fig:toposector} shows an
example of how the non-intersection constraint can prevent annihilation of two loops of
the same color.  Each sector corresponds to a
conserved charge in the transfer matrix; we will define these in 
section \ref{sec:transfer}. When building a quantum model based on this
classical model, each of these sectors will correspond to a ground
state of an appropriately defined Hamiltonian. We will discuss the
quantum model further in section \ref{sec:topological}.

Finally, when the model is defined on a torus, all noncontractible
loops must go around the same cycle. This cycle can be labeled as
$m\vec{c}_1+n\vec{c}_2$, where $m$ and $n$ are integers, and
$\vec{c}_1$ and $\vec{c}_2$ define the torus.  Most
topological sectors on the torus can be labeled by $S_c\times S_t$,
where $S_c$ is a non-trivial element of the free group defined above
with the additional requirement that products must be interpreted
cyclically, and $S_t$ is an element of $SL(2,\mathbb{Z})$, the group
of modular transformations of the torus. $SL(2,\mathbb{Z})$ is
generated by exchanging $m\leftrightarrow n$, and shifting $n\to n+1$.
Topological sectors not of this form are the trivial sector $S_c=1$
(i.e.\ no $S_t$), and sector with a single loop (i.e. $S_c=a,b$ or
$c$), where $S_t=\mathbb{Z}_2\times \mathbb{Z}_2$.

\section{Field-theory approaches}
\label{sec:ft}

A basic question about our model is if it is critical.  We can gain
insight into this question by studying the field theories valid near
two critical points which occur when by relaxing or increasing the
constraints.

Three decoupled Ising models are critical when $K$ is appropriately
tuned. In the continuum limit, the critical point can be described by
using conformal field theory \cite{Belavin84}.  One important thing
conformal field theory allows one to do is classify all the operators
of the theory. The Ising model has only two relevant
rotationally-invariant ones, the spin field, and the energy operator
$\epsilon$. Perturbing the critical point by the latter corresponds to
changing the temperature, i.e.\ taking $K$ away from $K_c$, so in the
lattice model we can identify
$$\epsilon\sim s_i s_j \sim d_{\langle i j\rangle}$$ so that
$d_{\langle i j\rangle}=-1$ corresponds to a domain wall between $i$
and $j$. A useful symmetry of the Ising model is Kramers-Wannier
duality, which shows the equivalence of high- and low-temperature
partition functions. In terms of the spins/fields, it takes $d\to -d$
and $\epsilon\to -\epsilon$.

To reach our model 3, one must perturb the (Ising$)^3$ critical point
to enforce the constraint ${\cal C}_j=0$ from (\ref{constraint3}).  By
construction, ${\cal C}_j=0$ when the domain walls through this
hexagon do not cross, and ${\cal C}_j=6$ when they cross. Thus
to reduce the weight of configurations where domain walls cross, we
add ${\cal C}_j$ to the energy with positive coefficient $\lambda$,
i.e.\
$$E= E_0 + \lambda \sum_j {\cal C}_j$$ where $E_0$ is the energy of
three decoupled Ising models. The constraint (\ref{constraint3}) is
enforced in the $\lambda\to\infty$ limit.  This perturbation is
clearly relevant, since it includes the energy operators $d_j$ in the
three individual models, and marginal terms which couple the two
models. A key fact to notice is that ${\cal C}_j$ is not invariant
under any of the dualities of the three Ising models.

An important result in two-dimensional statistical mechanics is the
existence of a ``$c$-theorem'' \cite{Zamo86}. The $c$-theorem says
that there is a function $c$ of the parameters of the theory satisfying
a very important property: it cannot increase under renormalization
group flows. Moreover, at a critical point its value is known from
conformal field theory -- it is a quantity called the central charge.
Thus if one starts at a known critical point and perturbs by a
relevant operator, the fact that $c$ must decrease means that either
the flow must end up at a critical point with a smaller value of $c$,
or at no critical point at all.  

The Ising critical point has $c=1/2$. Since our model is a relevant
perturbation of three Ising models, this implies that either our model
is not critical, or if it is critical, it should have $c\le 3/2$. 
The three-color
model (at infinite temperature) is critical, and has $c=2$\ 
\cite{Reshetikhin91,Read92,Kondev95}. Thus imposing constraint
\ref{constraint2} on the three-color model should move the model away
from the three-color critical point. This is in accord with the
numerics discussed below.

An obvious question is if our model is critical at some value of
$K$. While it is conceivable, it does not seem likely. One can cancel
the relevant piece $d_j$ of ${\cal C}_j$ by changing the temperature
of the three Ising models. This leaves the marginal terms quadratic in
$d_j$. These marginal terms can change the dimensions of operators, so
a fine-tuned model could be critical. However, since the constraint
${\cal C}_j=0$ violates dualities, this critical point is not likely
to be the (Ising$)^3$ one. This argument does not preclude a flow to a
critical point with a lesser central charge.

%The relation in the previous section to the zero-temperature Ising
%antiferromagnet and the honeycomb dimer model gives valuable
%information on such a flow. The honeycomb dimer model is critical, and
%has central charge $c=1$ \cite{kondevhenley}. 

One candidate for a flow is the hard-hexagon model. The critical point
in this model has $c=4/5$; it is in the same universality class as the
three-state Potts model \cite{Baxbook}. However, it occurs very far
from the hard-hexagon model of interest, for several reasons. First,
to get our model, one must allow configurations not present in the
hard-hexagon model.  Second, the latter's critical point occurs when
the weight $z$ per hard hexagon is $z=z_c\equiv
((1+\sqrt{5})/2)^5=11.09...$. To get model 3 at infinite temperature,
the configurations must all be of equal weight, i.e.\
$z=1$. Perturbing $z$ away from $z_c$ is relevant. It is not clear
whether allowing the additional configurations is relevant or not. It
is conceivable that the two perturbations could effectively cancel,
leaving one at the hard-hexagon critical point, but we have no
evidence for this.

There are no unitary critical points with $Z_3$ symmetry and $c<4/5$,
so our model cannot be critical with these central
charges. There are several with $4/5 < c < 3/2$, so it is conceivable
that it could be critical with these central charges, but we have
found no evidence for this.

%relevance of 3-color perturbation???

%%%%%%%%%%%%
\section{The transfer matrix}
\label{sec:transfer}

In this section we define the transfer matrix, and show that it
possesses some remarkable and unusual properties at infinite
temperature. We exploit these properties in the next section to solve
our model on a different lattice, and then in section
\ref{sec:numerics} to do numerics on very large systems.

Consider the model formulated in terms of non-crossing domain walls on
the links of the triangular lattice. Take the transfer matrix to act
perpendicular to one of the three axes. The transfer matrix $T$ acts
on the space of states on a zig-zag line; each link is labeled by an
index $i=1\dots 2L$, where $L$ is the number of hexagons across the
original lattice. We take the convention that links $i$ and $i+1$ meet
at a vertex when $i$ is odd. The degrees of freedom are the domain
walls on the links. We denote $w_i=1$ if there is a domain wall on
link $i$, and $w_i=0$ if there is not. The space of states is then of
dimension $2^{L}$.

The interactions are at the vertices of the lattice: the fact that
there are only 10 vertices must be enforced. It is most convenient to
write $T$ in the form
\begin{equation}
T=U{\cal T}U{\cal T}^{-1}
\label{TUTU}
\end{equation}
where $U$ moves you to the next zig-zag line, which has the property
that links $i$ and $i+1$ meet at a vertex when $i$ is even.
$U$ therefore imposes the weights at $L$ vertices. The
operator ${\cal T}$ is the translation operator, which shifts all the spins
by one site. The transfer matrix with periodic boundary conditions in
both directions is therefore
$Z= \hbox{tr}(T^M)$ 
for a lattice of $2LM$ hexagons. Since $T$ and $T{\cal T}^2$ have the
same eigenvectors, the eigenvectors of $T$ are the same
as those of $U{\cal T}$, and the eigenvalues are simply related.

Some interesting conservation laws follow immediately from the fact
that domain walls do not cross. The total number of domain walls must
be conserved mod 2, so $\sum w_i$ is conserved mod 2.  When $N$ is a
multiple of three, this conservation law is much more powerful: the
transfer matrix locally conserves the number of domain walls mod 2 on
{\em each} of the three sublattices. Namely, just as the sites can
be divided into three sublattices, the links can as well; these result
in the three types of the domain walls illustrated in figure
\ref{fig:3Ising}. The power of the non-crossing
constraint is that adjacent domain walls of different types cannot
change places or annihilate as the transfer matrix evolves the
system across the lattice. The
distinct sectors on the cylinder described at the end of section
\ref{sec:flip} are a consequence of this symmetry.

% Label the
%links of each type by $\alpha=0,1,2$, where $\alpha= i-3[i/3]$, where
%$[x]$ is the integer part of $x$. Then each configuration along the
%zig-zag line can be labeled by a string indicating which values of
%$\alpha$ the domain walls have, e.g.\ $0120110120$. This string is
%ordered with the walls; because of the periodic boundary conditions it
%should be interpreted cyclically. The only annihilation which can happen is
%that two adjacent labels of the same type (e.g.\ $11$) can
%annihilate. This results in a large (exponentially-increasing in $L$)
%number of different sectors preserved by the transfer matrix. These
%sectors are labeled by different valus of this string (up to cyclic
%permutations and removing identical adjacenct values). For example,
%the sector for the example given above can be labeled $1212$. We will
%see that the sector with the largest eigenvalue of $T$ is, not
%surprisingly theone with no label, i.e.\ with configurations that can
%annhilate to give no domain walls. 

This symmetry is already quite powerful. By studying $U$, we find even
more remarkable properties. $U$ commutes with $2L$ { local} 
symmetry generators, so the model with transfer matrix $U$ instead of
$T$ has a {\em gauge symmetry}.

There are two different types of local conservation laws. The first
one is easy to see.  Say two consecutive links meeting at a vertex are
both occupied, i.e.\ $w_{2j-1}w_{2j}=1$. Then examine the 10 vertices
in figure \ref{fig:10vertex2}, and take the transfer matrix to act in
the vertical direction. There is only one possible vertex where both
incoming links are covered, the last one drawn. This vertex has both
outgoing links covered as well. Thus acting with $U$ keeps
$w_{2j-1}w_{2j}=1$, while all other vertices have $w_{2j-1}w_{2j}=0$
before and after $U$ acts. Thus $Q_j\equiv w_{2j-1}w_{2j}$ is conserved by
$U$ for any integer $j$.

The second local conservation law is not as obvious.  It involves two
adjacent vertices connected by a horizontal link. This horizontal link
is of the same type as the links $2j-1$ and $2j+2$, so an incoming
domain wall on these links can turn by 120 degrees onto the horizontal
link.  This conservation law arises from the facts that there are no
allowed vertices which have just one or three walls touching them, and
that the number of walls of a given type is conserved {\em locally}
mod 2.  To illustrate this, first consider the case where links $2j-1$
and $2j+2$ are either both occupied, or both unoccupied. Computing $U$
requires summing over the two possibilities for the horizontal
link. When the horizontal link is unoccupied, the only allowed
contribution to $U$ is to leave the configuration unchanged. When the
horizontal link is occupied, the only allowed configuration is that
both-occupied annihilates into both-unoccupied, or vice
versa. Therefore $U$ here does not conserve $w_{2j-1}$ and $w_{2j+2}$
individually, but it does preserve the number of incoming lines mod 2.
Defining $R_j=(2w_{2j-1}-1)(2w_{2j+2}-1)$, here we have $R_j=1$ before and
after $U$ acts. When one of the two links $2j-1$ and $2j+2$ is
occupied and the other unoccupied, $R_j=-1$. In this case, when the
horizontal link is unoccupied, $U$ leaves the configuration unchanged,
and when the horizontal link is occupied, the two configurations
change place. Thus $R_j$ remains $-1$ before and after $U$ acts. Thus
$R_j$ is a local conserved quantity.  Note that $\prod_j R_j =(-1)^W$,
where $W=\sum_i w_i$ is the total number of walls (which is indeed
conserved mod $2$).

The quantities $Q_j$ and $R_j$ are not conserved in the full model,
because they do not commute with translation operator ${\cal T}$.
However, in the infinite-temperature case $K=0$, they do allow the
non-zero eigenvalues of the full transfer matrix $T$ to be found from
much smaller matrices. For example, we
show in section \ref{sec:numerics} how this gives the largest eigenvalue of
$T$ for $L=6$ from a 5-by-5 matrix, considerably smaller than the
$2^{12}\times 2^{12}$ transfer matrix obtained without exploiting any
symmetries!

The key simplification in the $K=0$ limit is that $U$ becomes a sum of
projection operators. Precisely, for each set of values of the $Q_j$
and $R_j$, define a matrix ${\cal P}(\{Q_j\},\{R_j\})$ acting on the
$2^{2L}$ states on the zig-zag line. The matrix elements ${\cal
P}(\{Q_j\},\{R_j\})_{ab}$ are defined to be 1 if both states $a$ and
$b$ have the charges $\{Q_j\},\{R_j\}$, and 0 if either or both do
not. Then the result is that
\begin{equation}
U=\sum{\cal P}(\{Q_j\},\{R_j\})\qquad\qquad\hbox{ when } K=0,
\label{UP}
\end{equation}
where the sum is over all possible values of $Q_j=0,1$ and $R_j=\pm 1$.
Note that not all values are 
possible: for example, if $Q_j=1$ and $Q_{j+1}=1$, then $R_j$ must
be $1$ as well. The decomposition (\ref{UP}) follows from an extension of the
arguments which led to $[U,Q_j]=[U,R_j]=0$. There we saw that each
initial state leads to an outgoing state with the same charges at most
once.  Since $K=0$, all allowed configurations have the same weight 1,
so every entry of $U$ must be $0$ or $1$. Moreover, by explicitly
examining all the possibilities for each set of four successive sites
$2j-1,2j,2j+1,2j+2$ and the horizontal links touching the two
vertices, it is easy to see that $U$ takes any initial state with a
given values of $Q_{j},Q_{j+1},R_j$ and $R_{j+1}$, to any final state with the
same values. Thus $U$ indeed is block diagonal, with each block
given by the operator ${\cal P}(\{Q_j\},\{R_j\})$.

Let us give an explicit example with $L=2$ and periodic boundary conditions.
We denote a state with domain walls at $i,j,k$ by $(i,j,k)$,
and the empty state as $()$.
Consider the sector which has
have the same conserved charges as the empty state, which are
$Q_1=Q_2=0$, and $R_1=R_2=1$. The other states which have these
charges are $(1,4),(2,3)$. It is then easy to check that on these
three states 
$$U={\cal P}(\{0,0\},\{1,1\})=\begin{pmatrix}
1&1&1\\
1&1&1\\
1&1&1
\end{pmatrix}
$$
There are two states in each of the other sectors with $Q_1=Q_2=0$.
When $R_1=-R_2=1$, the sector is comprised of $(2)$
and $(3)$, when  $R_1=-R_2=- 1$ it consists of $(1)$
and $(4)$, and when  $R_1=R_2=-1$, it consists of $(1,3)$ and
$(2,4)$.  Within any of these sectors,
$$U=\begin{pmatrix}
1&1\\
1&1
\end{pmatrix}\ .
$$
$U$ for each of the 7 states with $Q_1=1$ and/or $Q_2=1$ 
is diagonal: there is only one state in each sector. 

The crucial property of $U$ at $K=0$ is that it is proportional to a
projection operator. Namely, the product of two different projection
operators is zero, and each ${\cal P}$ ${\cal P}^2=n{\cal P}$,
where $n$ is the number of states in this sector. Each ${\cal P}$ has
only a single non-zero eigenvalue $n$, and the corresponding
eigenstate is the equal-amplitude sum over all states in the
sector. Thus most states in the Hilbert space are annihilated by $T$.
The eigenstates of $T$ with non-zero eigenvalues have an important
property, following from the fact that all states in the same sector
end up with the same coefficient after acting with $U$.  Since $U$ is
the last part of of $T$, the final state after acting with $T$ must
have the same property: all states with the same values of $\{Q_j\}$
and $\{ R_j\}$ have the same coefficient in the end. This means that
at $K=0$, all eigenstates of $T$ with non-zero eigenvalue must have
the same property as well!

We can therefore work in a space of states vastly reduced in size, by
keeping just one state in each sector. How to work out the explicit
transfer matrix in this reduced basis is explained in section
\ref{sec:numerics}. We emphasize that the $Q_j$ and $R_j$ are not
conserved charges for the full transfer matrix $T$, like they are for
$U$. The eigenstates of $T$ do not have definite values of the $Q_j$
and $R_j$, but are a sum over states with different values. Our result
here says that for eigenstates of $T$ with non-zero eigenvalues, all
states in a given sector must have the same coefficient. This is not a
symmetry, because the coefficients are not the same for eigenstates
with zero eigenvalue.

\section{The gauge-symmetric model}
\label{sec:gauge}

Since the matrix $U$ commutes with all the local symmetry generators,
using it as a transfer matrix results in a model with a gauge
symmetry. Because of the gauge symmetry, the resulting ``model $U$''
can be reduced to a one-dimensional model and solved exactly. In this
respect it is quite similar to the two-dimensional Ising gauge theory.
However, the solution of model $U$ has some very striking properties
of its own: the eigenvalues of the transfer matrix are given in terms
of Fibonacci numbers. We derive this here.

\begin{figure}[h]
\centerline{\includegraphics[width=.4\textwidth]{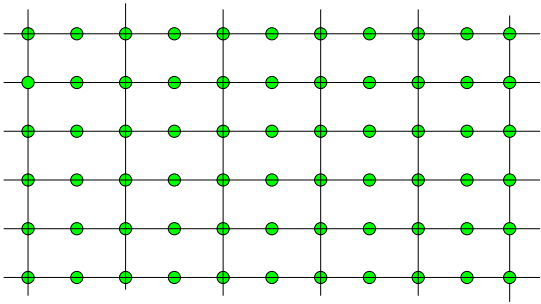}}
\caption{The lattice for model $U$.}
\label{fig:modelU}
\end{figure}

Model $U$ is the Ising model with
constraint (\ref{constraint1}) around each plaquette of the lattice
pictured in figure \ref{fig:modelU}. It is the square lattice, with an
extra site added to all the horizontal links. It is therefore not
rotationally invariant.
We find explicit expressions for the eigenvalues of $U$ in the limit
$K=0$, where we can exploit the fact that its transfer matrix $U$ can
be written as the sum (\ref{UP}). This means that the eigenstates are
the sum over all states in a given sector, and the corresponding
eigenvalue is the number of states in that sector. This turns out to
be an amusing combinatorial problem.

Let us consider the sector including the empty state, which has all
$Q_j=0$ and all $R_j=1$. Having $Q_j=w_{2j-1}w_{2j}=0$ means that the
links $2j-1$ and $2j$ are not both occupied by walls. Having $R_j=1$
means that either both of the links $2j-1$ and $2j+2$ are occupied,
are neither one is. The eigenvalue for ${\cal
P}(\{0,0,\dots\},\{1,1,\dots\})$ is then the number of states
$\Lambda_0$ satisfying these constraints. To count these, note that if
both links $2j-1$ and $2j+2$ are occupied, then links $2j$ and $2j+1$
must be unoccupied, in order to preserve $R_j=R_{j+1}=0$. But if these
latter two links are unoccupied, then links $2j-3$ and $2j+4$ must be
unoccupied as well, to keep $Q_{j-1}=Q_{j+1}=1$. This rule gives a way
of counting the configurations in this sector using one-dimensional
transfer matrix $V$, which propagates the system by two sites. Start
at one end. If $w_{1}=w_{4}=0$, then $w_{3}=w_{6}$ can be either $0$
or $1$. However, if $w_{1}=w_{4}=1$, then $w_{3}=w_{6}=0$. Iterating
this procedure along the whole line gives
\begin{eqnarray*}
\Lambda_0 &=& \hbox{tr}(V^L),\\
V&=&
\begin{pmatrix}
1&1\\
1&0
\end{pmatrix} 
\label{Vdef}
\end{eqnarray*}
The first row and column of $V$ correspond to unoccupied links, while
the second correspond to occupied ones.
It is simple to show by induction that
$$V^a =\begin{pmatrix}
F_{a+1}&F_a\\
F_{a}& F_{a-1}
\end{pmatrix}
$$
where $F_{a}$ is the $a$th Fibonacci number ($F_0=0$, $F_1=1$, and
$F_a= F_{a-1}+F_{a-2}$ for the rest). Thus
$$\Lambda_0 = F_{L+1} + F_{L-1}$$
which for large $L$ grows as $\tau^{L+1}$, where $\tau=(1+\sqrt{5})/2$
is the golden mean. 

Using a transfer matrix in one dimension makes it possible to write an
expression for all the eigenvalues. First consider the case with
all $R_j=1$ except $R_1=-1$, and all $Q_j=0$. If $w_1=1-w_4=0$, then
$w_3=1$, but if $w_1=1-w_4=1$, then $w_3$ can be either $0$ or
$1$. Thus the 1d transfer matrix for $j=1$ is $VA$, where
$$A=\begin{pmatrix}
0&1\\
1&0
\end{pmatrix}\ .
$$
The eigenvalue $\Lambda_1$ for the case where one of the $R_j$ is
flipped to $-1$ is therefore
$$\Lambda_1 = \hbox{tr}(AV^L) =2 F_L = \Lambda_0 - F_{L-3}.$$
In general, when a given $R_k=-1$, one simply inserts $A$ at the
$k$th site. Thus if $R_k=R_{k+a}=-1$ with all others remaining $1$, we
have eigenvalue
$$\Lambda_{a,L-a} = \hbox{tr}(AV^aAV^{L-a})\ .$$
By using various identities for Fibonacci numbers, one finds
$$\Lambda_{a,L-a}= \Lambda_0 - F_{L-a}F_a\ .$$

Letting some of the $Q_j$ be $1$ can be handled in a similar fashion. As
noted above, having $Q_k=1$ means that the walls on links $k-2$ and
$k+3$ automatically follow from knowing $R_{j-1}$ and $R_j$. This is
handled in the transfer-matrix formalism by inserting the matrix $VB$
at the site of every $Q_k=1$, where 
$$B=\begin{pmatrix}
0&0\\
0&1
\end{pmatrix}\ .
$$
Thus when $Q_k=1$ for some $k$ while
all other $Q_j=0$ and all $R_{j}=1$, we have eigenvalue
$$\hbox{tr}(BV^L)=F_{L-1}=\Lambda_0 - F_{L+1}$$
This eigenvalue is smaller than $\Lambda_1$ and $\Lambda_{a,L-a}$;
the eigenvalue $\Lambda_0$ is the largest, with
$\Lambda_1$ the next highest.

Continuing in this fashion, one obtains the general formula
\begin{eqnarray}
\Lambda(\{Q_j\},\{R_j\})&=& \hbox{tr}
\left( VX_L VX_{L-1} \dots VX_2 VX_1\right),
\label{general1}\\
X_j &=& A^{(1+R_j)/2} B^{1-Q_j}
\label{general2}
\end{eqnarray}

The conservation laws still hold when $K\ne 0$, so $U$ remains block
diagonal. However, the blocks are no longer projection operators, so
there is generically more than one non-zero eigenvalue per block. We
suspect however that the gauge symmetry makes it possible
the eigenvalues here in terms of
a one-dimensional transfer matrix like
(\ref{general1},\ref{general2}).

\section{Numerical results}
\label{sec:numerics}

To use exact diagonalization on the transfer matrix of the full model
at $K=0$,
we utilize the trick described in section \ref{sec:transfer} to reduce
its size.  This enables us to find its largest eigenvalue for cylinders
of up to $L=18$ hexagons (36 Ising sites).

%\begin{center}
%\begin{widetext}                    
\begin{table*}
\begin{tabular*}{5.5 in}{@{\extracolsep{\fill}} c   c  c c c c c}
Width ($\hexagon$)&\multicolumn{2}{c}{$S/\hexagon$}&
\multicolumn{2}{c}{$c_{\rm est}$}&\multicolumn{2}{c}{$f_{0,{\rm est}}$}\\[0.5ex]
\hline
&Model 3&Three-color&Model 3 & Three-color&Model 3 & Three-color\\
\hline\hline\\
3&0.4621&0.4621\\
6&0.3911&0.4028&1.880&1.569& 0.3674&0.3830\\
9&0.3771&0.3900&2.000&1.829&0.3659&0.3798\\
12&0.3722&0.3853&1.990&1.914&0.3660&0.3793\\
15&0.3700&&1.965&&0.3660\\
18&0.3688&&1.942&&0.3661\\
$\vdots$&$\vdots$&$\vdots$\\
 $\infty$&0.3661&0.3791&&1.99&0.3661&0.3791\\
 \hline
Theory &&0.379114&&2&&0.379114\\
\end{tabular*}
\caption{\label{table1}
Results of numerical transfer-matrix calculations on model 3
with periodic boundary conditions, compared to three-color model.
The estimated central charge $c_{\rm est}$ and bulk free energy
$f_{0,{\rm est}}$ are obtained for $n$ hexagons by fitting the
entropy values for $n$ and $n-3$ hexagons to equation (\ref{cftfit}).
Extrapolations of entropy per site to the infinite system fit the
last three points to $c_0 + c_2 L^{-2} + c_4 L^{-4}$.}
%\end{ruledtabular}
\end{table*}
%\end{widetext}
%\end{center}

Each state in this new space is labeled by the values $\{Q_j\}$ and
$\{R_j\}$, which for short we call $B$.  After $U{\cal T}$ acts,
giving every element in the same block the same coefficient, we label
the blocks $B'$. To work out the transfer matrix in this new basis,
first one needs to list all the states in a given block $B$.  Pick one
and act with ${\cal T}$, i.e.\ just shift the whole thing over by one
site. Compute the new values of $\{Q_j\}$ and $\{R_j\}$ after the
shift, or equivalently, compute $\widetilde{Q}_j=w_{2j}w_{2j+1}$ and
$\widetilde{R}_j=(2w_{2j}-1)(2w_{2j+3}-1)$, which we collectively
label $\widetilde{B}$.  The block $B'$ reached from acting with
$U{\cal T}$ on this element $B$ is then labeled by $\{Q_j\}$ and
$\{R_j\}$, where $Q_j=\widetilde{Q}_j$ and $R_j=\widetilde{R}_j$ for
all $j$. One does this for each element in the block $B$: work out
$\widetilde B$ and then $B'$ for each, and then increase the element
${\cal R}_{B'B}$ by one. Going through all the blocks gives the
reduced transfer matrix ${\cal R}$.

Since the
eigenvectors of $U{\cal T}$ are the same as those of $T$, and the
eigenvalues are simply related, we focus on this. To give an example,
for $L=2$, we have
$$U{\cal T}=\begin{pmatrix}
1&0&0&1&1\\
1&0&0&1&1\\
1&0&0&1&1\\
0&0&1&0&0\\
0&1&0&0&0
\end{pmatrix}\ ,
$$ which has eigenvalues $2,-1,0,0,0$. Note that it is not
symmetric. There are three blocks here. The block $B=1$ has three
states $(),(14),(23)$, the block $B=2$ has just $(12)$, and the block
$B=3$ has just $(34)$.  Upon acting with ${\cal T}$, $()$ goes to the
block $()$, which means $U{\cal T}$ takes it to all the members of
this block. Thus we increase ${\cal R}_{11}$ by 1.  Acting with ${\cal
T}$ on $(14)$ takes it to $(12)$, so we increase ${\cal R}_{21}$ by
1. Acting with ${\cal T}$ on $(23)$ gives $(34)$, so ${\cal
R}_{31}=1$. Doing this for the other two blocks gives
$${\cal R}=\begin{pmatrix}
1&1&1\\
1&0&0\\
1&0&0
\end{pmatrix}
$$
This has eigenvalues $2,-1,0$ as we want, and is symmetric -- it just
lost some zero eigenvalues.

The ground-state sector for $L=3$ is even easier. There are 4 states
in the block $1$: $(),(14),(25),(36)$. These all go to the same block
under $U{\cal T}$, so the reduced transfer matrix is simply a number:
$4$. This is indeed the largest eigenvalue here. For higher $L$, the
size of ${\cal R}$ still increases exponentially, but not as
quickly. To give an example of how much this reduces the size of the
matrix, exploiting translation invariance and
parity as well means that the largest eigenvalue of $U{\cal T}$ 
for $L=6$ is the same as that of the 5-by-5 matrix
$$
\begin{pmatrix}
4&6&6&0&2\\
2&6&4&2&0\\
1&2&2&0&0\\
0&1&0&0&0\\
1&0&0&0&0
\end{pmatrix}
$$ In this and all the examples we have examined, ${\cal R}$ is upper
{\em left} triangular.

We have used this reduced transfer matrix for the domain-wall loop
representation in numerical simulations of transfer matrices with
width up to 18 hexagons in model 1 (36 Ising
variables). The resulting largest eigenvalues for widths that are
multiples of 3 are shown in table 1.  The entropy per hexagon
converges to a number right in the middle of the upper (from the
three-color model) and lower (from the hard-hexagon model) bounds
given in (\ref{bounds}).

Expanding the largest eigenvalue in a power series in $1/L$ gives
additional valuable information. When the system is at a conformally
invariant critical point, the subleading piece is universal and
proportional to the central charge \cite{BCNA}, which must obey $c\ge
1/2$ in any system with positive Boltzmann weights. If the system is not
critical, this piece should fall off to zero as
$L\to\infty$.  The precise formula for our case is
\begin{equation}
f = \frac{\log \Lambda}{L} 
= f_0 + \frac{\pi c}{ 6} \frac{\sqrt{3}}{ 2} \frac{1}{L^2} + \dots.
\label{cftfit}
\end{equation}
Here $L$ is the width in hexagons, $\Lambda$ is the largest eigenvalue
of $U{\cal T}$, and the geometrical factor $\sqrt{3}/2$ results
from the ratio between the width and length of the transfer matrix
step.  The resulting estimates of central charge for our model do not
converge even at the largest system sizes, while for the three-color
model, extrapolation from smaller sizes gives a central charge
consistent with the expected value $c=2$ ~\cite{Reshetikhin91,Read92,Kondev95}.

The conclusion of this transfer-matrix study is that our model is most
likely not described by the $c=2$ critical theory of the three-color
model, even though large system sizes are required to see the
difference. Since the central charge does not seem to be converging to
anything, the numerical results are in harmony with the field-theory
arguments of section \ref{sec:ft} in suggesting that our model is non-critical
(i.e., has a finite correlation length). We cannot categorically rule
out that it is critical with $c<2$, but have no evidence for this
scenario.

\section{Further directions: the quantum theory}
\label{sec:topological}

In this paper we discussed constrained classical lattice models. By
imposing some simple constraints on Ising spins, we found a variety of
intriguing properties. In particular, we showed that the space of
states on the sphere is connected under local moves, and that on
surfaces with non-contractible cycles, different sectors can be
labeled by loop configurations. We also presented substantial (if not
conclusive) evidence that the model is not critical.

In the introduction, we mentioned a quantum
motivation for studying classical lattice models with constraints. 
The results of this paper imply that our model has the right
characteristics to yield a quantum model with a topological phase,
with the added intriguing possibility that the excitations have
non-abelian statistics. We therefore 
will conclude this paper with a discussion of the quantum model in
more detail.

The connection between quantum and classical models 
comes from a trick due to Rokhsar and Kivelson
\cite{Rokhsar88}. Let the basis elements for the Hilbert space for the
quantum model consist of configurations in the two-dimensional
classical lattice model. Then one can construct a quantum Hamiltonian
acting on these states with a ground state consisting of a {\em
superposition} of these states, with each term having an amplitude
corresponding to its weight in the classical model. Correlators in the
ground state of the quantum model are then related to the correlators
in the classical model.
 
In the quantum model, one need not impose the constraints directly on
the Hilbert space, but rather one can add an energy penalty for
configurations which violate the constraint. The ground state then
contains only configurations satisfying the constraint. Violating the
constraint locally then corresponds to a quasiparticle
excitation. Thus for a given classical model, one can obtain very
different quantum models depending on which defects are allowed and
which are not. 

Let us make this explicit in terms of model 1. Here we can take the
Hilbert space to be comprised of two-state Ising variables on the
sites of the honeycomb lattice. One simply imposes an energy penalty
on configurations violating constraint (\ref{constraint1}), i.e.\ for
each hexagon $h$ one includes $M_h^2$ in the Hamiltonian. The
off-diagonal terms in Hamiltonian are given by flip we defined in
section \ref{sec:flip}, as displayed in figure
\ref{fig:flippable}. The trick of Rokhsar and Kivelson is to add a
potential which, combined with the flip, is a projector. Namely, we
add a potential term which counts the number of flippable
hexagons. Then the Hamiltonian is
$$H_1=\sum_h \left({\cal N}_h -{\cal F}_h + (M_h)^2\right)$$ where ${\cal
F}_h$ is the flip and ${\cal N}_h$ is the number of flippable
plaquettes. It is simple to write $H_1$ explicitly in terms of Ising
spins, but the expression is quite unwieldy. The lowest eigenvalue of
$H_1$ is zero, since it is the sum of projectors and a positive
diagonal term. The ground state on the sphere is unique, and
consists of the equal-amplitude sum over all
configurations satisfying constraint (\ref{constraint1}). On surfaces
with contractible cycles, the local flip cannot change the sectors
described above. There will be a ground state for each of these
sectors, consisting of the equal-amplitude sum of all the
configurations in the sector satisfying the constraint.

The reason for the interest in quantum models of this type is that
they often have topological order. Topological order means that
there is no local order parameter with a non-vanishing expectation
value, but only non-local ones. By using the Rokhsar-Kivelson trick,
it was demonstrated that a quantum eight-vertex model \cite{Kitaev97}
and a quantum dimer model on the triangular lattice \cite{Moessner01}
indeed have topological order. The quantum model with Hamiltonian $H_1$ 
indeed should have topological order, since the number of ground
states depends on the genus of the surface, a telltale sign.
One reason why models with topological order are interesting is that
they can lead to excitations with fractional statistics.   Non-topological solid and superfluid
quantum phases near the Rokhsar-Kivelson point corresponding to our model,
and possible unconventional phase transitions, are discussed using a
two-component quantum height model in Ref.\ ~\onlinecite{ashvinunpub}.

\begin{figure}[h]
\centerline{\includegraphics[width=.45\textwidth]{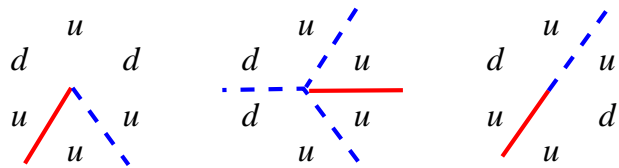}}
\caption{Defects occurring in model 1 when $M_h=2$.}
\label{fig:defects}
\end{figure}

The excitations in the quantum model with Hamiltonian $H_1$ correspond
to hexagons with different numbers of up and down spins. The specific
$M_h$ we chose means the lowest-energy defects have $M_h=\pm 2$. These are
illustrated in figure \ref{fig:defects}.
In the loop language, they correspond to joining loops of different
types. With the Hamiltonian $H_1$, these defects have no dynamics, but
one can of course add terms allowing them to move. 

Changing the potential to favor other kinds of defects gives different
theories.  Allowing just $M_h=\pm 4$ defects gives a variation on the
``odd'' Ising gauge theory described in depth in \cite{MSF}; these
defects correspond to allowing loops (the domain walls of model 3) to
cross. Allowing just $M_h=\pm 6$ defects gives a quantum version of
the three-color model, i.e.\ the defects are the 11th vertex shown in
figure \ref{fig:11thvertex}.  Note however that even though the ground
states of all three models we have introduced are identical, their
local defects can be quite different.  The defects just described are
nonlocal in the color representation of model 2, because the three
colors become rotated upon circling, i.e., each bond no longer has a uniquely defined color\cite{moorelee}. Likewise, another
kind of defect we could introduce would be to treat the {\em loops} in
model 3 as the degrees of freedom for the quantum model. Then we can allow
defects to correspond to loops with ends (like the end of a flux
tube in gauge theory).

The defects we have discussed have an important property: although the
energy $M_h^2$ associated with them is local, they are attached to
zero-energy defect lines, which can only end in another defect. For
example, a $M_h=\pm 2$ defect has two types of domain walls attached,
which must eventually end in another defect. This property makes it
likely that the corresponding quasiparticles have fractional
statistics, because when particles are exchanged, they must pass
through these defect lines.

The $M_h=\pm 2$ defects illustrated in figure \ref{fig:defects} are
particularly intriguing. Since the model has an $S_3$ symmetry under
exchange of the Ising models, these defects can be classified in
representations of this non-abelian symmetry.  (Note also that the
larger symmetry generated by the global conserved quantities of the
classical transfer matrix is non-abelian as well.) This makes it possible
that a suitable choice of Hamiltonian will result in non-abelian
braiding of the excitations, a topic of great current interest because
of potential application to topological quantum computation \cite{tqc}. 

These arguments make it likely that one can realize a phase with
topological order using our model as a starting point. To prove this,
more work needs to be done. One needs to prove the quasiparticles are
deconfined, i.e.\ that lines connecting the defects have no energy
per unit length in the quantum theory. This does seem very plausible,
given that $M_h$ is non-zero only at the location of the defect.  In the
three-color model, these defects have binding free energy that scales
as a power-law~\cite{moorelee}, which is
critical between confinement and deconfinement.  A
related question is proving that the ground state of the Hamiltonian
contains macroscopically long loops even in the continuum limit; many
examples are known of lattice loop models where the average loop
length (in terms of the lattice spacing) remains finite. Also, the
Hamiltonian $H_1$ does not allow the defects and defect lines to cross
through each other, making it impossible to understand the fractional
statistics precisely.

We leave these very interesting open questions for future study.

\bigskip
We thank Ashvin Vishwanath for useful conversations.  This work was supported by the NSF under grant
DMR-0412956 (P.F.) and DMR-0238760 (J.E.M. and C.X).

\appendix
\section{Relations to other models}
\label{sec:relations}

Valuable intuition and information can be gained
by relating our model to two well-studied models, the
triangular-lattice Ising antiferromagnet and the hard hexagon model.
Our model can be found by {\em relaxing} constraints on these two. 
Since adding constraints reduces the entropy, the maps
described in this section give lower bounds on the entropy of our
model. Moreover, both have critical points different from that
of the three-color model.

The Ising antiferromagnet on the triangular lattice is one of the
classic examples of geometrically frustrated
magnetism~\cite{andersonfazekas}. At zero temperature in the classical
model, each fundamental triangle contains either two up spins and one
down, or two down spins and one up. To avoid confusing these Ising
spins with the earlier ones, we label them as $h_i=\pm 1/2$, so the
zero-temperature constraint is that the sum of spins around every
fundamental triangle is $\sum_\triangle h_i = \pm 1/2$.  By drawing
each frustrated bond as a dimer on the dual lattice, this model is
identical to the close-packed hard-core dimer model on the honeycomb
lattice, which is known to be critical \cite{Wu}.

We now consider a model with the same constraint around each triangle,
but where the degrees of freedom can take {\it any} half-integer value
$\pm 1/2, \pm 3/2, \ldots$, not just $\pm 1/2$ as in the Ising
antiferromagnet. We call this a ``height'', and prove here that this
model is equivalent to ours. The Ising spin $\sigma_i = \pm 1$ of
our model 1 is defined on the sites of dual honeycomb lattice by
\begin{equation}
\sigma_i = 2 (-1)^i\sum_\triangle h_i = \pm 1,
\label{sigmah}
\end{equation}
where even and odd $i$ are the sites on the two equivalent sublattices
of the honeycomb lattice. The constraint (\ref{constraint1}) follows
automatically from this definition.

\begin{figure}[h]
\centerline{\includegraphics[width=.2\textwidth]{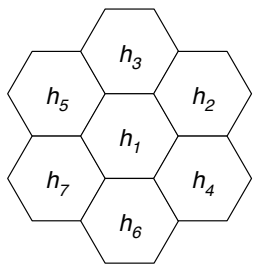}}
\caption{The labels used in the text to establish the equivalence between model 3 and a generalization of the triangular lattice Ising antiferromagnet.}
\label{fig:dualhex}
\end{figure}

Any height configuration obeying (\ref{sigmah}) therefore defines a
configuration in model 1.  To finish the proof of equivalence, we now
show that each Ising spin configuration satisfying constraint
(\ref{constraint1}) generates, up to two arbitrarily specified
half-integers, a unique configuration of heights. Fix a configuration
of Ising spins on the honeycomb lattice.  Pick two adjacent sites on
the dual triangular lattice, and assign them arbitrary half-integer
heights $h_1$, and $h_2$. If one knows two of the heights around a
triangle, and the value of the Ising spin at the center of the
triangle, then (\ref{sigmah}) determines the third height
uniquely. Consider the sites illustrated in figure \ref{fig:dualhex}. 
Applying (\ref{sigmah}) to the two triangles containing both $h_1$ and
$h_2$ gives the heights $h_3$ and $h_4$. Applying
(\ref{sigmah}) again to the triangles involving $(h_1,h_3)$ and
$(h_1,h_4)$ gives two more heights $h_5$ and $h_6$. There is now
another height $h_7$ which belongs to a triangle involving $(h_1,h_5)$
as well as to a triangle with $(h_1,h_6)$.  Again applying
(\ref{sigmah}) to either of these triangles gives $h_7$; because of
the constraint (\ref{constraint1}) it is determined uniquely. This
therefore determines all the heights on the six triangles involving
$h_1$. Repeating this process for the triangles around the heights
$h_2\dots h_7$ then determines the heights on another concentric
ring. In this fashion all the heights follow from the Ising-spin
configuration.  The constraint (\ref{constraint1}) ensures that these
are unique, up to the two original choices of $h_1$ and $h_2$.  The
indeterminacy of these two half-integers can be understood simply by
noting that, if three integers $a,b,c$ are added to the height
variables globally on the three sub-lattices of the triangular
lattice, then as long as $a+b+c=0$, the resulting Ising spin
configuration on the honeycomb lattice is unchanged.

Our model is obtained by relaxing a constraint on the zero-temperature
Ising antiferromagnet, so it provides a lower bound on the entropy of
our model. The honeycomb-dimer model equivalent to the former has an
entropy of $.323\dots$ per hexagon \cite{Kastelyn,Wu}. A slightly
better lower bound can be obtained by relating our model to another
interesting model, the hard hexagon model.

\begin{figure}[h]
\centerline{\includegraphics[width=.35\textwidth]{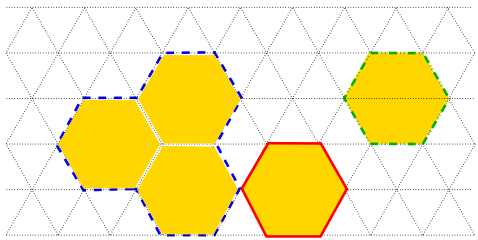}}
\caption{A typical configuration in the hard hexagon model. }
\label{fig:hardhex}
\end{figure}
The hard hexagon model is defined by placing particles
on the sites of the triangular lattice, so that no two particles are
adjacent or on the same site. Each particle can equivalently be viewed
as a ``hard'' hexagon with the length of a link: the
restriction that particles cannot be placed on adjacent sites means
that the hexagons may not overlap \cite{Baxbook}. A typical
configuration is drawn in figure \ref{fig:hardhex}.
The relation between model 3 and hard hexagons comes by drawing
lines surrounding any clusters of hexagons, as shown in figure
\ref{fig:hardhex}. Each of these loops corresponds to a domain wall in
one of the three Ising models on the three triangular sublattices. By
construction, these loops do not cross, although they can touch. Thus
each configuration in the hard-hexagon model corresponds to one in
model 3. The converse is not true: there are configurations in
model 3 not in the hard hexagon model. In model 3, one can have
domain-wall loops inside of other loops, as long as they do not cross. 
If there
is a domain-wall loop of one Ising model inside that of another, this
corresponds in the hard hexagon model to placing a hexagon on top of
others. This is forbidden there.

Both the hard hexagon model and the three-color model without
constraint (\ref{constraint2}) are integrable. In both cases, one can
compute the asymptotic behavior of the number of configurations as the
number of sites gets large \cite{Baxter70,Baxbook}. Since our model
has more configurations than the hard hexagon model
and less than the three-color model, this gives lower and upper bounds
on the entropy $S$ in this limit:
\begin{equation}
0.3332 < \frac{S}{N} <  0.3791\ ,
\label{bounds}
\end{equation}
where $N$ is the number of sites on the triangular lattice in model 3
(the number of hexagons in the honeycomb lattice in model 1).  Our
numerics discussed in section \ref{sec:numerics} give
$S/N=0.3661\dots$, consistent with these bounds.

%\end{document}

\end{document}